\documentclass[aps,pre,twocolumn,showpacs,showkeys]{revtex4}
\usepackage{graphicx}
\usepackage{amsmath}
\usepackage{amsfonts}
\usepackage[english]{babel}

\newcommand{\ef}[1]{\, #1}     

\newcommand{\eval}[1]{\left\langle {#1} \right\rangle}

\newcommand{\leval}[1]{\langle {#1} \rangle}

\newcommand{\tr}{\mathrm{tr}}
\newcommand{\diag}{\mathrm{diag}}

\newcommand{\vett}[1]
 {{\bf #1}}

\newcommand{\Ebar}{\bar{E}}

\begin{document}

\title{$O(n)$ vector model at $n=-1$, $-2$
       on random planar lattices:\\
       a direct combinatorial derivation}

\author{Sergio Caracciolo}
\email{Sergio.Caracciolo@mi.infn.it}
\author{Andrea Sportiello}
\email{Andrea.Sportiello@mi.infn.it}
\affiliation{Universit\`a degli Studi di Milano -- Dip.~di Fisica and INFN,\\
via Celoria 16, I-20133 Milano, 
Italy}

\date{\today}

\begin{abstract}
The $O(n)$ vector model with logarithmic action
on a lattice of coordination 3 is related to a gas of self-avoiding loops on the
lattice. This formulation allows for analytical continuation in $n$:
critical behaviour is found in the
real interval $[-2,2]$.
The solution of the model on random planar lattices, recovered by random matrices,  also involves an analytic
continuation in the number $n$ of auxiliary matrices.
Here we show that, in the two cases $n=-1$, $-2$,
a combinatorial reformulation of the loop gas problem allows to achieve
the random matrix solution with no need of
this analytical continuation. 
\end{abstract}



\maketitle

\section{Introduction}

Because of  conformal symmetry, universality in
$2$-dimensional statistical systems on a regular lattice is a
well-understood topic~\cite{difrancConf}. To perform
 a further average over the ensemble of random
planar lattices, in some cases, provides even deeper results, because
of the still larger symmetry (the discrete version of general
covariance)~\cite{David,Jurg,Kazakov}.

The generating function of
statistical configurations over random lattices can be written as the Feynman expansion
of a proper action: the replacement of real or complex bosonic fields
with $N \times N$ symmetric- or hermitean-matrix fields allows to
count lattices of genus $g$ with a weight proportional to $N^{-g}$,
and a large-$N$ limit, achieved via steepest descent or continuous
approximation of matrix spectra, gives the restriction to
planar lattices (see~\cite{difranc} for a recent pedagogical introduction).

Since the pioneering works on Ising and Potts Models
\cite{kazakov_ising}, almost all 
interesting statistical
2-dimensional problems have been studied also in their 
variant on random planar lattices. The specific case of $O(n)$ vector model with logarithmic
action on coordination-3 lattices, introduced by Nienhuis on the
regular hexagonal lattice \cite{nienhuis}, has been solved in the
random case by Kostov \cite{kostovMPL89}, and further studied in
particular by~\cite{Gaudin,eynardzinnjustin}.

For a given lattice $\Lambda$, with $V$ vertices, the $O(n)$ partition
function is here defined as
\begin{equation}
\label{eq.Z.fixlat}
Z_n(\Lambda, \beta)=
\int \prod_{i=1}^V 
\mathrm{d}^n\vett{s}_i  \frac{2\,\delta(|\vett{s}_i^2|-1)}{\Omega_n}
\prod_{\eval{ij}} (1+ \beta n \, \vett{s}_i \cdot \vett{s}_j)
\end{equation}
 where $\Omega_{n}$ is the surface of the unit sphere $S^{n-1}$ in $n$-dimensions,
 so that
 \begin{equation}
 \label{eq.spheresurf}
 \Omega_n=\frac{2 \pi^{\frac{n}{2}}}
 { \Gamma \left(\frac{n}{2}\right)},
 \end{equation}
 and it is introduced in order to have $Z_n(\Lambda, 0)=1$.
 
 $O(n)$-invariance implies
\begin{align}
\label{eq.vevs}
\eval{1} &=1 \ef;
&
\leval{s^{(\alpha)}_{i} s^{(\beta)}_{i} } &= \frac{1}{n} \delta_{\alpha, \beta}
\ef;
\\
\leval{s^{(\alpha)}_{i} } &= 0 \ef;
&
\leval{s^{(\alpha)}_{i} s^{(\beta)}_{i} s^{(\gamma)}_{i} } &= 0
\ef.
\end{align}
Note that higher-order momenta are not relevant, as they do not appear
in the polynomial expansion of (\ref{eq.Z.fixlat}), this being a
specific feature of the logarithmic action and the choice of the coordination number.

In each monomial of the
expansion, each edge $(i,j)$ can contribute either with a factor 1, or with a
factor $\beta n s^{(\alpha)}_i s^{(\alpha)}_j$, with 
$\alpha \in \{1, \ldots, n\}$. In this last case, we
interpret an edge $(i,j)$ as marked, and labeled by $\alpha$.

In order to have a non-vanishing
contribution after integration, each vertex must have an even number
of marked edges of each species stemming from it, 
thus, as the coordination is 3, the only contributing configurations are
the ones of self-avoiding unoriented loops, coloured with an index
$\alpha$. Summing over possible loop labelings leads to a factor $n$ per
loop, and we are left with a sum over configurations $L$ of 
self-avoiding unlabeled unoriented loops 
\begin{equation}
\label{eq.Zloops}
Z_n(\Lambda, \beta)= \sum_L n^{k(L)} \beta^{|L|}
\ef,
\end{equation}
where $k(L)$ is the number of loops and $|L|$ is the number of edges
in $L$. In this form of loop gas, the problem has a natural analytic extension
to complex values of $n$. Note that we can alternatively sum over oriented loop
configurations $L^*$
\begin{equation}
\label{eq.Zloops2}
Z_n(\Lambda, \beta)= \sum_{L^*} 
\left( \frac{n}{2} \right)^{k(L^*)} \beta^{|L^*|}
\ef.
\end{equation}
A typical configuration over a random planar lattice looks like the
one in figure~1. 
The model shows a nontrivial critical behaviour in the range $-2 \leq n \leq 2$,
the cases $n=1$ and $n=2$ corresponding respectively to Ising and XY Model. 
In the case $n\to 0$ only one loop survives and when $\beta=\infty$ this model deals with Hamiltonian cycles~\cite{Kristjansen}.
The case $n=-1$ can also be seen as a particular $q \to 0$ limit of the $q$-state Potts
Model~\cite{Alanpottsrev}. Recently, it has also been connected with the generating function of
forests on the graph~\cite{noi}.

In this letter we will show that, in the two cases $n=-1$, $-2$, 
a direct combinatorial reformulation of the loop gas problem  allows to achieve
the random matrix solution with no need of analytical continuation in $n$.

\begin{figure}[t]
\label{fig_1}
\begin{center}
\setlength{\unitlength}{35pt}
\begin{picture}(5,4)
\put(0,0){\includegraphics[scale=1.4, bb=0 0 125 100, clip=true]{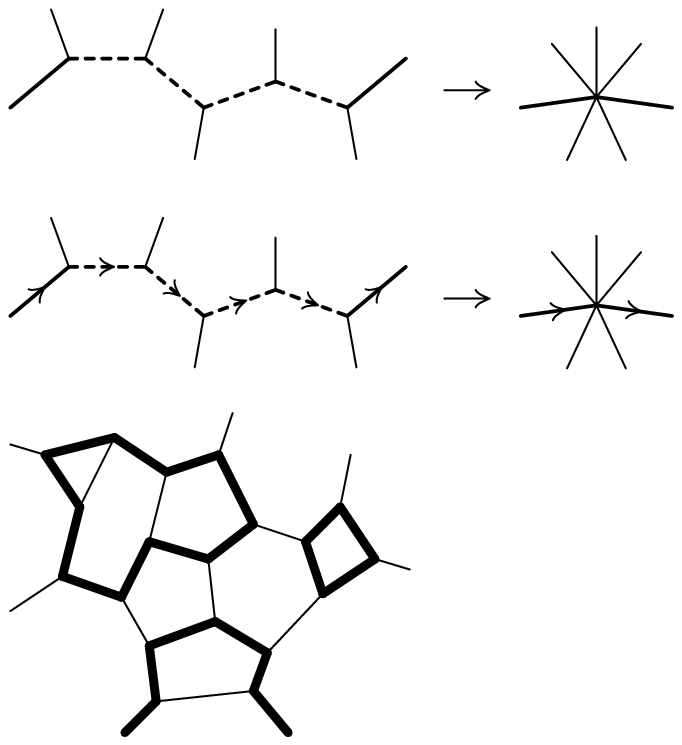}}
\end{picture}
\caption{A portion of a typical configuration 
of loop gas on a random 3-graph.}
\end{center}   
\end{figure}

\section{Sketch of the Random Matrix Solution}

Consider the set $\mathcal{L}$ of planar graphs $\Lambda$ with
coordination 3. We are interested in the double-series generating function
\begin{equation}
\label{eq.Z.RM}
Z_n(g, \beta)=
\sum_{\Lambda \in \mathcal{L}}
\frac{g^{|\Lambda|}}{|\mathrm{Aut}(\Lambda)|}
Z_n(\Lambda, \beta)
\ef,
\end{equation}
where $|\Lambda|$ is the number of vertices in the lattice.
When dealing with the $O(n)$ model, $n+1$ matrix fields
$M$, $E_1$, \ldots, $E_n$ are introduced~\cite{kostovMPL89,Gaudin,eynardzinnjustin}. 
The quadratic part of the action is
\begin{equation}
\label{eq.S0.k}
\mathcal{S}_0=
\tr \left[
\frac{1}{2} \left( M^2 + E_1^2 + \cdots + E_n^2 \right)
\right]
\ef;
\end{equation}
while the interaction part is
\begin{equation}
\label{eq.SI.k}
\mathcal{S}_I= 
\tr \left[
-\frac{g}{3} M^3 
- g \beta \left( E_1^2 + \cdots + E_n^2 \right)M
\right]
\ef;
\end{equation}
and the partition function (\ref{eq.Z.RM}) is given by
\begin{equation}
Z_n(g, \beta)=\lim_{N \to \infty} \frac{1}{N^2}
\ln \int \mathcal{D}_N(M,\left\{ E_{\alpha} \right\})
e^{-N(\mathcal{S}_0 + \mathcal{S}_I)}
\ef.
\end{equation}
Notice that auxiliary matrices $E_{\alpha}$ only appear quadratically,
although with a non-trivial tensorial form in matrix indices.
At this point one can apply a change of variables. Say
$M=O \vec{\lambda} O^{-1}$, with $O \in O(N)$ and 
$\vec{\lambda}=\diag(\lambda_1, \ldots, \lambda_N)$. Change variables
from $E_{\alpha}$ to $E'_{\alpha}= O^{-1} E_{\alpha} O$. Integration
over $M$ can be replaced by integration over $\vec{\lambda}$:
\begin{equation}
\int
\mathcal{D}_N(M)
=
\int
\mathrm{d}^N\vec{\lambda}\, \Delta^2(\vec{\lambda})
\ef,
\end{equation}
with the Jacobian given by the square Vandermonde determinant
\begin{equation}
\Delta^2(\vec{\lambda})
=
\prod_{i \neq j}
|\lambda_i - \lambda_j|
\ef.
\end{equation}
We are left with
\begin{multline}
Z_n(g, \beta)=\lim_{N \to \infty} \frac{1}{N^2}
\ln 
\int
\mathrm{d}^N\vec{\lambda}\, \Delta^2(\vec{\lambda})
\\
\int \mathcal{D}_N(\{E_{\alpha}\}) \,
e^{-N(\mathcal{S}'_0 + \mathcal{S}'_I)}
\ef,
\end{multline}
with
\begin{align}
\label{eq.S0.kbis}
\mathcal{S}'_0
&= \frac{1}{2} \sum_i \lambda_i^2 + \frac{1}{2}
\sum_{i,j} \sum_{\alpha=1}^n (E_{\alpha})_{ij} (E_{\alpha})_{ji}
\ef;
\\
\label{eq.SI.kbis}
\mathcal{S}'_I
&= 
-\frac{g}{3}
\sum_i \lambda_i^3
+\beta g \sum_{i,j} \sum_{\alpha=1}^n
(E_{\alpha})_{ij} (E_{\alpha})_{ji} \lambda_i
\ef.
\end{align}
We integrate over variables $(E_{\alpha})_{ij}$, and, up to an irrelevant constant, we are left with
\begin{multline}
\label{eq.kostov}
Z_n(g, \beta)=\lim_{N \to \infty} \frac{1}{N^2}
\ln 
\int
\mathrm{d}^N\vec{\lambda}\, \Delta^2(\vec{\lambda})
\\
\exp \Big[
-N \Big(\frac{1}{2} \sum_i \lambda_i^2 -
\frac{g}{3} \sum_i \lambda_i^3 \Big) \Big]\,
\prod_{i,j} (1-\beta g (\lambda_i + \lambda_j))^{-\frac{n}{2}}
\end{multline}
This result is, for example, the starting point for all the following analysis in Kostov article \cite{kostovMPL89}.

\section{Alternative combinatorics for loop gas}
\label{sec.loop}

For a given lattice $\Lambda$ of coordination 3, call $\mathcal{P}$ (respectively $\mathcal{P}^*$)
the set of configurations $P$ of unoriented (respectively oriented) self-avoiding open paths,
and $|P|$ the number of edges in $P$.
Say that $P \subseteq L$ if all the edges in $P$ are edges of the loop $L$, and
$P^* \subseteq L^*$ if also orientation is preserved.

Introduce the two partition functions
\begin{subequations}
\label{eq.Zloop}
\begin{align}
Z(\Lambda,\beta,\gamma)
&=
\sum_P \beta^{|P|} \sum_{L \supseteq P} \gamma^{|L|-|P|}
\\
Z^*(\Lambda,\beta,\gamma)
&=
\sum_{P^*} \beta^{|P^*|} \sum_{L^* \supseteq P^*} \gamma^{|L^*|-|P^*|}
\end{align}
\end{subequations}
Say $L$ (resp. $L^{*}$) is composed by loops $\{ \ell_{k} \}$, of
$|\ell_k|$ edges, then the quantities (\ref{eq.Zloop}) are equal
respectively to
\begin{subequations}
\label{eq.Zloopbis}
\begin{align}
Z(\Lambda,\beta,\gamma)
&=
\sum_{L} \prod_k \Big( (\gamma+\beta)^{|\ell_k|} - \beta^{|\ell_k|} \Big)
\\
Z^*(\Lambda,\beta,\gamma)
&=
\sum_{L^{*}} \prod_k \Big( 2(\gamma+\beta)^{|\ell_k|} -
2\beta^{|\ell_k|} \Big)
\end{align}
\end{subequations}
that is, each edge of $L$  (resp. $L^{*}$) can be in $P$  (resp. $P^{*}$) or not,
independently, but for the global constraint that not all the edges
of a given loop are in $P$  (resp. $P^{*}$), which accounts for the subtractions in
formulas (\ref{eq.Zloopbis}). Remark that
\begin{align}
Z(\Lambda,\beta,-\beta) &= Z_{-1}(\Lambda,\beta)
\\
Z^*(\Lambda,\beta,-\beta) &= Z_{-2}(\Lambda,\beta)
\end{align}

\section{Alternative Random Matrix Formulation of the problem}
\label{secIV}

Now we want to restate the combinatorial theory described in section
\ref{sec.loop}, averaged over the set of random planar 3-graphs, as a
random-matrix integral. We start from the case $n=-1$:
\begin{equation}
Z_{-1}(g,\beta)=
\sum_{\Lambda \in \mathcal{L}}
\frac{g^{|\Lambda|}}{|\mathrm{Aut}(\Lambda)|}
Z(\Lambda, \beta, -\beta)
\ef.
\end{equation}
Apparently, this theory requires three symmetric matrices: a matrix $M$ for
edges not in $L$, a matrix $E$ for edges in $L$, but not in $P$, and a
matrix $F$ for edges in $P$.
This is not the case: the constraint that $F$-edges do not make loops
is not local, and cannot be implemented by a local action. It is
necessary to build effective vertices of arbitrary degree, 
containing the connected components of $P$, in order to satisfy the
constraint. So, as all $F$-edges are treated explicitly in
vertex-combinatorics, they do not play a role in Feynman
diagrammatic, and both the integration and the action only depend over
matrices $M$ and $E$, the quadratic part of the action being
\begin{equation}
\label{eq.S0.my}
\mathcal{S}_0
=
\tr \left[
\frac{1}{2} \left( M^2 + E^2 \right)
\right]
\ef;
\end{equation}
As in the previous treatment,
we have the pure $M$- and $E$-edge vertices
\begin{equation}
\label{eq.SI.my1}
\mathcal{S}^{(1)}_I = 
\tr 
\bigg[ - \frac{g}{3} M^3 + g \beta E^2 M \bigg]
\ef,
\end{equation}
where the factor $-\beta$ in the second term accounts for the two external
$E$-legs.
This corresponds to vertices of $\Lambda$
which are not adjacent to edges of $P$.
In the other case, say that a path of $P$ has $p\geq 2$ vertices (in
order to be non-empty): the corresponding vertex
of the action must be built shrinking all the edges stemming from the
path into a single point. The weight of the effective vertex is
$-(\beta g)^p$.
It has $p$ external $M$-edges, and two $E$-edges, in a
certain cyclic sequence.
Summing over all possible orderings gives a combinatorics
\begin{equation}
-\frac{(\beta g)^p}{2} \sum_{p'=0}^p \binom{p}{p'}
E M^{p'} E M^{p-p'}
\end{equation}
and thus a further term in the interaction part of the action is
\begin{equation}
\label{eq.SI.my2}
\mathcal{S}^{(2)}_I
= 
\tr 
\sum_{p=2}^{\infty} 
\frac{(\beta g)^p}{2} \sum_{p'=0}^p \binom{p}{p'}
E M^{p'} E M^{p-p'}
\ef.
\end{equation}
Again the change of variables $M=O \vec{\lambda} O^{-1}$,
$E'=O^{-1} E O$ allows to perform first an angular integration, then the
gaussian integration of auxiliary variables $E_{ij}$. The various
terms in the action combine to give
\begin{equation}
\begin{split}
\mathcal{S}' &=
\frac{1}{2} \sum_i \lambda_i^2 
- \frac{g}{3} \sum_i \lambda_i^3 
\\
& \quad +
\frac{1}{2}
\sum_{i,j}
\sum_{p=0}^{\infty} 
(\beta g)^p \sum_{p'=0}^p \binom{p}{p'}
E_{ij} \lambda_j^{p'} E_{ji} \lambda_i^{p-p'}
\\
&= 
\frac{1}{2} \sum_i \lambda_i^2 
- \frac{g}{3} \sum_i \lambda_i^3 
+ \frac{1}{2}
\sum_{i,j} 
\frac{E_{ij} E_{ji} }
{1- \beta g (\lambda_i + \lambda_j)}
\ef,
\end{split}
\end{equation}
and a final Gaussian integration gives the quantity
(\ref{eq.kostov}), for $n=-1$.

\begin{figure}[t]
\label{fig_2}
\begin{center}
\setlength{\unitlength}{30pt}
\begin{picture}(8,4.6)
\put(0,0){\includegraphics[scale=1.2, bb=0 100 200 240, 
clip=true]{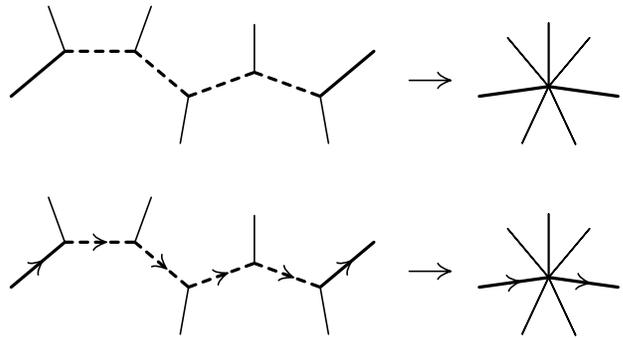}}
\end{picture}
\caption{Contraction of the effective vertex, for the theory at $n=-1$
(top) and $n=-2$ (bottom). Thin, thick-solid and thick-dashed 
lines correspond respectively to $M$-, $E$- and $F$-edges.}
\end{center}
\end{figure}


The case $n=-2$ is very similar. We have to calculate
\begin{equation}
Z_{-2}(g,\beta)=
\sum_{\Lambda \in \mathcal{L}}
\frac{g^{|\Lambda|}}{|\mathrm{Aut}(\Lambda)|}
Z^*(\Lambda, \beta, -\beta)
\ef.
\end{equation}
Again we have to work only with two matrices, $M$ and $E$, but now, as
edges in $L$ are oriented, $E$ is an hermitean-matrix field, the
quadratic part of the action being
\begin{equation}
\label{eq.S0.my.n2}
\mathcal{S}_0
=
\tr \left[
\frac{1}{2} M^2 + \Ebar E
\right]
\ef;
\end{equation}
while the pure $M$- and $E$-edge vertices give
\begin{equation}
\label{eq.SI.my1.n2}
\mathcal{S}^{(1)}_I = 
\tr 
\bigg[ - \frac{g}{3} M^3 + g \beta \Ebar E M \bigg]
\ef.
\end{equation}
For the remaining part of the action,
say that an oriented path of $P$ has $p\geq 2$ vertices.
The weight of the effective vertex is still
$-(\beta g)^p$, but now
it has $p$ external $M$-edges, one $\Ebar$-edge and one $E$-edge, 
in a certain cyclic sequence.
Summing over all possible orderings gives a slightly different
combinatorics
\begin{equation}
-(\beta g)^p \sum_{p'=0}^p \binom{p}{p'}
\Ebar M^{p'} E M^{p-p'}
\end{equation}
and thus a further term in the interaction part of the action is
\begin{equation}
\label{eq.SI.my2.n2}
\mathcal{S}^{(2)}_I
= 
\tr
\sum_{p=2}^{\infty} 
(\beta g)^p \sum_{p'=0}^p \binom{p}{p'}
\Ebar M^{p'} E M^{p-p'}
\ef.
\end{equation}
Again perform the change of variables,
then angular integration:
the various terms in the action combine to give
\begin{equation}
\begin{split}
\mathcal{S}' &=
\frac{1}{2} \sum_i \lambda_i^2 
- \frac{g}{3} \sum_i \lambda_i^3 
\\
& \quad +
\sum_{i,j}
\sum_{p=0}^{\infty} 
(\beta g)^p \sum_{p'=0}^p \binom{p}{p'}
\Ebar_{ij} \lambda_j^{p'} E_{ji} \lambda_i^{p-p'}
\\
&= 
\frac{1}{2} \sum_i \lambda_i^2 
- \frac{g}{3} \sum_i \lambda_i^3 
+
\sum_{i,j} 
\frac{\Ebar_{ij} E_{ji} }
{1- \beta g (\lambda_i + \lambda_j)}
\ef,
\end{split}
\end{equation}
and a final Gaussian integration gives the quantity
(\ref{eq.kostov}), for $n=-2$.

\section{A Final Remark}

Calculations of section \ref{secIV} can be performed 
also for generic $\gamma$ different from $ - \beta$. We have 
\begin{gather}
\label{eq.gammabeta}
\begin{split}
Z(g, \beta, \gamma) &=\lim_{N \to \infty} \frac{1}{N^2}
\ln 
\int
\mathrm{d}^N\vec{\lambda}\, \Delta^2(\vec{\lambda})
\\
& \quad \cdot \, \exp \Big[
-N \Big(\frac{1}{2} \sum_i \lambda_i^2 -
\frac{g}{3} \sum_i \lambda_i^3 \Big) \Big]\,
\\
& \quad \cdot \,\prod_{i,j} 
\left(
\frac{1-\beta g (\lambda_i + \lambda_j)}
{1-(\gamma+\beta) g (\lambda_i + \lambda_j)}
\right)^{\frac{x}{2}}
\ef;
\end{split}
\end{gather}
with $x=1$ for unoriented 
 (resp.~$x=2$ for oriented) loops.
More generally, if the weight associated to a loop
of length $\ell$ is $\sum_B \beta_B^{\ell} - \sum_F \beta_F^{\ell}$,
the whole procedure can be repeated,
the last factor in the expression for the random lattice partition
function being
\begin{equation}
\prod_{i,j} 
\left(
\frac{\prod_F (1-\beta_F g (\lambda_i + \lambda_j))}
{\prod_B (1-\beta_B g (\lambda_i + \lambda_j))}
\right)^{\frac{x}{2}}
\ef.
\end{equation}
We have denoted the positive weights $\beta_{B}$ because we have seen how the bosonic matrices
$E$'s naturally give rise to the jacobian in the denominator. Weights $-\beta_{F}$ could be obtained 
either introducing fermionic matrix fields,
or by the combinatorial method discussed in this paper. Cancellations between numerator and denominator for
equal values of $\beta$ are a manifestation of Parisi-Sourlas supersymmetry~\cite{PS}.


\begin{acknowledgments}
We thank L.G.A.~Molinari for fruitful discussions.
\end{acknowledgments}


\end{document}